\documentclass[twocolumn,showpacs,showkeys,amsmath,amssymb,nofootinbib,prd,floatfix]{revtex4}
\usepackage{graphicx}
\usepackage{times}

\renewcommand{\d}{\partial}
\newcommand{\const}{{\rm const}}
\begin{document}

\title{Dark matter dynamics in Galactic center}
\author{Eugene Vasiliev and Maxim Zelnikov}
\email{eugvas@lpi.ru, zelnikov@lpi.ru} 
\affiliation{Lebedev Physical Institute, Leninsky prospekt 53, Moscow, Russia}

\date{February 29, 2008}

\begin{abstract}
The evolution of dark matter in central areas of galaxies is considered (the Milky Way is taken as an example). 
It is driven by scattering off of dark matter particles by bulge stars, their absorption by the supermassive 
black hole and self-annihilation. 
This process is described by diffusion equation in the phase space of energy and angular momentum. 
The equation was integrated for several different models of initial dark matter distribution and using various 
assumptions about the dynamical factors. 
It turns out that because the Milky Way center is rather dynamically old ($\sim 4$ relaxation times $t_r$), 
the difference in initial conditions almost vanishes. 
The density attains a nearly universal profile, and the $\gamma$-ray flux from dark matter annihilation lies 
in rather narrow range, which enables more robust determination of the dark matter parameters. 
By present time the mass of dark matter inside the black hole sphere of influence ($r<2$~pc) has been reduced 
approximately twice, mostly because of heating by stars.
It is shown that the dynamics of dark matter for $t\gtrsim t_r$ is determined mainly by stars outside 
the sphere of influence.
\end{abstract}



\maketitle

\section{Introduction}

It is widely accepted that the largest fraction of matter in the Universe is the dark matter (DM) \citep{a0110404}. 
Its nature is still unclear, but most likely it consists of yet undiscovered cold (non-relativistic) particles, 
which have very low interaction cross-sections with each other and with baryonic matter \citep*{h0404175}. 
So the main physical mechanism governing the evolution of dark matter is gravitation. 

The dark matter is responsible for formation of large-scale structure of the Universe, as well as 
extended DM haloes of galaxies \citep{a0504097}. However, the very centers of galaxies (except low surface brightness 
galaxies) are dominated by baryons, which form galactic bulges. Their influence on DM haloes consists of at least 
two effects. 
Firstly, as the baryons cool and settle down in the center of a potential well created by DM, they change the common 
gravitational potential, which leads to compression of DM halo. This effect is called baryonic compression and 
increases DM density several times \citep{Blum, a0406247, a0507589, a0601669}. 
On the other side, dark matter particles are scattered off by
bulge stars and captured by supermassive black holes (SMBHs), which reduces their density in the inner parts of bulges 
\citep*{a0306490, a0307524, a0501555}. Additionally, a non-zero annihilation cross-section of DM particles also leads to 
decrease of DM density in the very centers of galaxies and gives possibility of indirect detection of DM through its 
annihilation radiation \citep{Zeld, a9906391, h0506204, a0504422}. 
All these issues are addressed in the present paper.

The paper is organized as follows. 
In the second section we describe previous investigations on this subject and explain motivation of this study. 
In the third section we present our class of initial DM halo models, which is somewhat broader than in previous 
papers. Then the kinetic equation governing the evolution of DM is derived, along with its coefficients and 
boundary conditions.
The fourth section is devoted to solution of the above equation, given for a series of initial halo models with 
inclusion of different physical processes and approximations. This helps us to clarify validity of some commonly used 
assumptions, and determine parameters on which the evolution depends mostly. Predictions for annihilation radiation are 
also given in this section.
Finally, the conclusions are presented.

\section{Overview of dark matter evolution in galactic centers}  \label{sec_overview}

The question about dark matter content in central areas of galaxies has long been studied. First of all, it depends on 
initial structure of DM halo. There are two principal classes of methods of halo structure modelling. Analytical 
methods generally consider the nonlinear stage of collapse of initial overdensity region. These include kinetic theory 
of phase mixing \citep{GZ} or variants of self-similar infall models \citep{a0008217, a0405491, a0506571}. In general, 
they predict a power-law density profile in the center of a DM halo, 
\begin{equation}  \label{rho_d}
\rho_d \propto r^{-\gamma_d} \,,
\end{equation}
where the power-law index $\gamma_d$ is between 1 and 2. A significant drawback of this approach is very limited 
ability to handle mergers, especially major ones. Alternatively, numerical simulations of dissipationless {\it N}--body 
gravitational interaction have been largely used in recent years. They also generally predict density profiles that 
have power-law cusps in central areas (e.g. Navarro, Frenk, White (NFW) profile \citep*{a9611107} with $\gamma_d=1$ or 
Moore profile \citep{a9903164} with $\gamma_d=1.5$). Other models were proposed in which the density slope tends to 
zero at $r \to 0$ (Einasto profiles \citep{a0311231, a0502515} or Burkert profiles \citep{a0506538}).  
There is still no consensus on the question whether dark matter halos are cuspy or cored in the centers, which means 
that central density profiles obtained in simulations may equally good be described with different fitting formulae, 
although the inferred densities in the central parsecs would vary orders of magnitude. 
The largest simulations of a galaxy-sized halo (e.g. \citep{a0611370}) provide measurements of density no closer 
to the galaxy center than $3\times 10^{-3} r_{vir} \simeq 10^3$~pc, which is still many orders of magnitude greater 
than the distances relevant to annihilation and scattering.

However, there are other processes that modify dark matter distribution in the galactic centers after halo formation. 
First of all, it is now well established that virtually all galaxies harbour supermassive black holes (SMBHs) in 
their centers \citep{a0411247}, whose masses $M_{bh}$ range from $10^6 M_\odot$ to more than $10^9 M_\odot$.
A black hole (BH) immersed in a stellar cluster (or bulge as a whole) modifies gravitational potential at distances 
smaller than the gravitational influence radius of the black hole $r_h$, defined as containing stellar mass equal to 
$2 M_{bh}$. 
If the black hole grew adiabatically, the surrounding matter (stars or DM) forms a ``spike'' with power-law index 
$\gamma' = (9-2\gamma)/(4-\gamma)$, $\gamma$ being the power-law index of density profile before black hole formation
\citep{a9906391}. 

The above argument applies for spherically symmetric evolution. If the black hole formed off-center and then spiralled 
in, the resulting profile should be shallower \citep*{a0101481}. Additionally, in the hierarchical merging 
scenario, binary SMBHs should be ubiquitous. A binary SMBH ejects matter from $r\lesssim r_h$ due to slingshot effect, 
and after coalescence the stellar and dark matter density around it is reduced \citep{a0201376}. 
However, it is unlikely that a major merger took place in the recent $10^{10}$~yr of Milky Way history \citep{a0501555}.

If the DM consists of self-annihilating particles, the product of annihilation cross-section $\sigma_a$ and 
relative velocity $v_r$ being weakly dependent on $v_r$, then the maximum density after a period of time $t$ 
is believed to be set at the so-called annihilation plateau: 
\begin{equation}  \label{rho_a}
\rho_a = m_p/ \langle \sigma_a v_r \rangle t 
\end{equation}
($m_p$ is the particle mass) \citep{h0404175}. However, this result is exact only for circular particle orbits.
If we consider arbitrary velocity anisotropy $\beta=1-\sigma_t^2/2\sigma_r^2$ (particulary, isotropic velocity 
corresponding to $\beta=0$), 
then we obtain a weak cusp instead of a plateau: its density varies as $r^{-(\beta+1/2)}$ and approximately equals 
corresponding plateau density $\rho_a$ at the radius where plateau itself should transform into initial density 
profile. This issue is explained in detail in \citep{a07073334}.

Finally, a very important process that modifies the dark matter distribution is the gravitational scattering of DM 
particles by stars in the vicinity of the SMBH and capture of particles by the SMBH. The detailed analysis of 
this process is the main subject of the present investigation. 

The stellar distribution around a massive black hole was much studied in 1970s \citep{BW, LS, CK}. 
The usual method includes representation of the distribution function (DF) $f$ in terms of energy $E$ and angular 
momentum $L$ per unit mass (for spherically symmetric problem these are the only two independent variables)
and consideration of diffusion in the $\{E, L\}$ phase space. 
In these early studies it was shown that diffusion along $L$ axis leads to formation of logarithmic dependence of $f$ 
on $L$, so that there is continuous flux of stars towards low values of $L$, where they are captured by black hole. 
The diffusion along $E$ axis drives the DF to a quasi-stationary state $f \propto |E|^{1/4}$, when the loss of stars 
is balanced by a inward flux from higher values of $E$ (outside the BH sphere of influence). The corresponding density 
profile $\rho \propto r^{-7/4}$ is called Bahcall-Wolf cusp, and is indeed observed in the center of our Galaxy 
\citep{a0305423, a0306130, a0703178}, though the density slope is somewhat lower. It is important that the 
stationary solution applies only to systems older than two-body relaxation time \citep{BT}
\begin{equation}  \label{t_relax}
t_r = \frac{0.34 \sigma^3}{G^2 m_\star \rho_\star \ln\Lambda} \:,
\end{equation}
where $\sigma$ is 1-d velocity dispersion, $\rho_\star$ -- stellar density, $\ln\Lambda \approx 15$ -- Coulomb 
logarithm. It appears that in the center of our Galaxy $t_r \sim 2.5\times 10^9 \mbox{yr} < t_{Hubble}$ and weakly 
depends on radius at $r<r_h$. (We adopt $M_{bh} = 3\times 10^6 M_\odot$ for consistency with other papers, though 
recent estimates are somewhat higher \citep{a0502129}. This yields $r_h \approx 2$~pc).

Dark matter dynamics is governed by the same mechanism of relaxation on stars, but due to negligible particle mass 
the effect of dynamic friction is unimportant, and resulting density profile should be $\rho_d \propto r^{-3/2}$. 
This simple argument was considered in \citep{a0308385}. 
A more elaborate treatment involves time-dependent solution of diffusion equation. This was first done in 
\citep{a0306490, a0307524} for a constrained initial DM distribution function using only diffusion 
along angular momentum. In \citep{a0311594} the problem was treated as one-dimensional diffusion for energy with 
inclusion of loss-term due to angular momentum diffusion and capture by BH, which was taken from \citep{CK}.
In subsequent papers \citep{a0501555, a0504422} the effect of self-annihilations was also included. These studies 
have led to the conclusion that the scattering off by stars drives DM distribution towards a quasi-stationary state
with diminishing amplitude of density profile due to combined effect of heating by stars and capture by the SMBH.
This density profile is regenerated within one $t_r$ even after a cusp has been destroyed by a binary SMBH 
\citep*{a0610425}.

We want to draw attention to several important issues that were not thoroughly investigated in previous studies.
\renewcommand{\theenumi}{\roman{enumi}}
\begin{enumerate}
\item The relaxation time actually depends on angular momentum $L$, as will be shown later. For small values of $L$ 
the diffusion along energy goes faster, so that the two-dimensional problem cannot be reduced to one-dimensional.
\item The black hole mass might have grown in time, which moves up the radius of BH influence and changes relaxation 
timescale inside this radius.
\item The initial DM distribution function need not be isotropic, i.e. might depend on angular momentum.
\end{enumerate}
All these issues are examined in the present paper. Additionally, the following features are discussed:
\begin{enumerate}
\item A broad class of initial distribution functions is considered. 
This enables direct comparison between different studies that took different initial conditions, and helps to 
disentangle the effect of initial conditions from other effects investigated.
\item The diffusion is considered both inside and outside the SMBH sphere of influence.
Values of diffusion coefficients are different in these two domains.
\item Analytical approximations of evolution of DM density profile and distribution function are made for limiting 
cases of $t\ll t_r$ and $t \gg t_r$. They are in good agreement with the solution of exact equation.
\item The two-dimensional diffusion equation is solved directly. The results are compared to the simplified case of 
one-dimensional treatment, to validate the possibility of such simplification.
\item The effect of self-annihilations is included as well. The proportion of captured, evaporated and annihilated 
dark matter mass is calculated.
\end{enumerate}

Having stated these important goals, we proceed to the investigation itself.

\section{Problem definition}

\subsection{Initial dark matter halo models}  \label{sec_initmodels}

Plenty of models have been suggested for DM haloes.
As we are interested in dark matter dynamics in galactic centers, we shall consider only internal part of halo, which 
can be reasonably approximated by a power-law density profile (\ref{rho_d}). We operate with a distribution function 
$f(E,L)$ so we need to transform $\rho_d$ to $f(E,L)$. This transformation is essentially non-unique -- one can obtain 
different DFs for the same density profile, which differ in dependence on angular momentum $L$. 
It is convenient to choose a factorized family of models $f(E,L) = f_1(E)\,f_2(R)$, where we have changed $L$ 
for a dimensionless scaled quantity $R \in [0..1]$:
\begin{equation}
R=L^2/L_c^2(E)\,,
\end{equation}$L_c$ being angular momentum of a circular orbit with given energy. (This has been suggested in
\citep{CK}). In the case of power-law density $f_1(E)$ should also be power-law \citep{BT}.

We consider two family of DFs, which differ in dependence on $L$:
\begin{subequations}
\begin{eqnarray}  \label{models}
\mbox{Model A:\!\!\!} && f(E,R)=f_0\,E^{1/2}\,\delta(R-R_0) \;,\\
\mbox{Model B:\!\!\!} && f(E,R)=f_0\,E^{1/2-(1-\beta)(4-\gamma_d)/(2-\gamma_d)}\, R^{-\beta}.\quad\;\;
\end{eqnarray}
\end{subequations}
Model A has been proposed in \citep{GZ} as a result of analytical treatment of DM halo collapse and phase mixing. The 
quantity $R_0$ is related to orbital eccentricities of particles and is small, $R_0 \ll 1$. (Further discussion can be 
found in \citep{a0601669}). 
Model B is the simplest generalization of isotropic distribution for the case of arbitrary radial velocity anisotropy 
$\beta$, considered in \citep{a0511686}. Here $\beta=1-\sigma_t^2/2\sigma_r^2$ is Binney's anisotropy parameter 
\citep{BT}: systems with $\beta=0$ are isotropic, with $0<\beta<1$ -- have radially biased velocities. 
For realistic DM haloes, $\beta$ in the center is zero or slightly positive \citep{a0405491}, as indicated also by 
a density slope--anisotropy relation proposed in \citep{a0411473}.

Concerning the density slope $\gamma_d$, we shall adopt values $\gamma_d=1$ (NFW profile \citep{a9611107}), 
$\gamma_d=1.5$ (Moore profile \citep{a9903164}) and $\gamma_d=0.25$
\footnote{This model approximates the inner 10 pc of the Einasto profile in \citet{a0504422}. Since it is much more 
difficult to incorporate profiles with variable density slope into our calculations, we have chosen a replacement for 
their ``standard halo model'' (SHM), that has a very mild cusp with average density inside 2 pc equal to that of SHM.}
for the model B   and $\gamma_d=12/7$ for model A \citep{GZ}. We use NFW or Moore profiles for most of our variants 
of calculation, since these are the most frequently used in the literature.

These density profiles refer to pure dark matter haloes. 
The central part of a halo is likely compressed in the process of galaxy formation and baryonic contraction
\citep{a0406247, a0601669}, which increases the slope of density profile as well as its normalization.
To account for adiabatic contraction, we transform $f(E,L)$ to $f(I,L)$, where $I$ is radial action -- a quantity 
that is conserved during slow change of potential. Hence $f(I,L)$ is conserved during compression. 
As the initial condition for diffusion equation we take $f'(E',L)$ converted back from $f(I,L)$ using now the potential 
of stars in the bulge and the SMBH. (In what follows we omit primes in $f'$ and $E'$).

All calculations are performed for the case of Milky Way.

The normalization of density profile is taken from the condition that $\rho_d(r_\odot) = \rho_\odot$, where 
$r_\odot = 8$~kpc is the distance to Galactic center, and $\rho_\odot=0.3 \mbox{ GeV/cm}^3$ \citep{a0501555}
\footnote{Except for model B4, which is normalized to match SHM of \cite{a0504422}, with density 40 $M_\odot/\mbox{pc}^3$ 
at 1 pc before contraction.}. 
We summarize initial model data in Table \ref{table_init_models}.

\begin{table}
\caption{Initial halo models used in calculations. $\gamma_d$ is the slope of density profile in the center before 
baryonic compression, $\beta$ is the velocity anisotropy parameter, $\rho_h$ and $\rho_h'$ are density values at
$r=r_h=2$~pc before and after compression, in $M_\odot/\mbox{pc}^3$. }  \label{table_init_models}
\tabcolsep=5pt
\begin{tabular}{lllll}
Model & $\gamma_d$ & $\beta$ & $\rho_h$ & $\rho_h'$ \\
\hline
A1 & 1.7 & 0.6 &$10^4$       &$2.3\times 10^4$\\
B1 & 1.5 & 0   &$2\times 10^3$&$1.3\times 10^4$\\
B2 & 1.5 & 0.5 &$2\times 10^3$&$1\times 10^4$\\
B3 & 1.0 & 0   & 30          &$3.4\times 10^3$\\
B4 & 0.25 & 0  & 30          &$1.8\times 10^3$\\
\hline
\end{tabular}
\end{table}

The density profile of bulge stars is taken in the form
\begin{equation}  \label{rho_star}
\rho_\star(r) = \rho_{\star,1} \left(\frac{r}{r_1}\right)^{-\gamma_\star}\;,\quad
\gamma_\star = \left\{ \begin{array}{rcl}1.5 &,& r<r_1 \\ 2 &,& r>r_1 \end{array} \right.
\end{equation}
Here $r_1=2$~pc, $\rho_{\star,1}=6\times 10^4\, M_\odot/\mbox{pc}^3$. Observations suggest similar 
broken power-law density profile with the same exponents, though with somewhat different values of $r_1$ and 
$\rho_{\star,1}$ \citep{a0305423} . The bulge outside SMBH radius of influence is therefore taken to be isothermal, 
with 1-D velocity dispersion $\sigma=80$~km/s  (this is true for inner 10--20 pc, see Fig.9 in \citep{a0203468}).
Relaxation time (\ref{t_relax}) inside $\sim 0.2r_h$ is independent of radius and equals $2.6\times 10^9$~yr. 
The velocity distribution of stars is assumed to be isotropic.

\subsection{Diffusion equation and its coefficents}

Gravitational scattering of DM particles by bulge stars leads to their diffusion in the $\{E, L\}$ phase space, 
which can be described by orbit-averaged Fokker-Planck equation \citep{BT}:
\begin{equation}  \label{diff_eq}
\frac{\d f}{\d t} = {\cal G}^{-1} \frac{\d}{\d \xi_\alpha}\left( {\cal G} 
\left[ D_{\alpha\beta} \frac{\d f}{\d \xi_\beta} - D_\alpha f \right] \right) - S_{ann}[f]\;,
\end{equation}
where $\xi_\alpha$ are phase-space variables (usually energy $E$ and angular momentum $L$, given per unit mass), 
$\cal G$ is the Jacobian, $D_\alpha$ and $D_{\alpha\beta}$ are drift and diffusion coefficients, respectively. 
(Given that DM particles have negligible mass, they are not subjected to dynamic friction and hence $D_\alpha=0$).
$S_{ann}$ is additional loss term due to particle annihilation.
We follow \citet{CK} in changing $L$ for dimensionless $R=L^2/L_c^2$, and also change $E$ for dimensionless variable 
$Q$ defined so that ${\cal G}=1$ over the whole region of interest (both the bulge, where $E>0$, and the 
SMBH region of influence, where $E<0$); $Q\to 0$ corresponds to particles very close to the SMBH with $E\to -\infty$,
$Q \sim 1$ separates the two regions. 
It can be easily shown that the quantity 
\begin{equation}  \label{Q_def}
Q = \frac{L_c^2(E)\, I_{r,0}(E)}{3 (\sigma r_h)^3 }
\end{equation}
satisfies the requirement (here $I_{r,0}(E) = I_r(E,0)$ is the radial action of orbit with $L=0$ and given $E$.
It can be shown that $I_r(E,R) \approx I_{r,0}(E)\,(1-\sqrt{R})$ \citep{a9906391, a0306490}).
The asymptotic expressions for $Q$ are the following:
\begin{subequations}
\begin{eqnarray}
Q=\frac{1}{24} \left( -E/2\sigma^2 \right)^{-3/2}       \quad   \;,&& E<0 \;,\\
Q=\frac{2}{3\rm e \sqrt{\pi}} \exp\left( 3E/2\sigma^2 \right)   \;,&& E>0 \;.
\end{eqnarray}
\end{subequations}

From the definition of $Q$ it follows that $Q$ is invariant in spherically-symmetric ($L=\const$) and 
adiabatic ($I=\const$) evolution, so that the distribution function in terms of $\{Q,R\}$ is conserved in the process 
of baryonic contraction and formation of the black hole. The initial power-law density profile corresponds to 
the following form of the distribution function:
\begin{equation}  \label{f_init_QR}
f_0(Q) = \tilde f(R) Q^{-(6-\gamma_d)/3(4-\gamma_d)} \,,
\end{equation}
where the dependence on $R$ is the same as in Eq.(\ref{models}).

The diffusion coefficients were derived in analytical form for the two limiting cases:
the bulge ($r\gtrsim r_h$, or $E>0$), and the SMBH region 
of influence, which differ in gravitational potential and stellar distribution function (details can be found in 
\citep{a0307524}). The expressions for $D_{\alpha\beta}$ in the SMBH region of influence are given below:
\begin{subequations}  \label{diff_coefs}
\begin{eqnarray}
D_{QQ,c} &=& H\,F_c\,\frac{36}{5}\, Q^2 \left(\frac{1}{\sqrt{R}}-\frac{1}{8}\right) \;,\\
D_{QR,c} &=& H\,F_c\,\frac{24}{5}\, Q\,R\left(1-\frac{1}{\sqrt{R}}\right) \;,\\
D_{RR,c} &=& H\,F_c\,R\,\frac{29+16\sqrt{R}-45R}{5} \;; \\
H        &=& \frac{16}3 \pi^2\,G^2\,m_\star\,\ln\Lambda   \;; \nonumber\\
F_c      &=& \frac{3\sqrt{2}}{32\pi^2\, G\sigma r_h^2} \;. \nonumber
\end{eqnarray}
\end{subequations}
$F_c$ is the star distribution function, which is independent of energy for $\gamma_\star=3/2$ in Coulomb potential;
$\ln\Lambda \approx 15$ is the Coulomb logarithm. (These coefficients may also be derived using expressions in 
\citep{CK} for $\gamma_\star=3/2$).

The interesting feature is that $D_{QQ}$ increases with decreasing $R$, which means that a particle with small angular 
momentum acquires energy faster (as its orbital pericenter lies in the higher density region). Usually this dependence 
of energy diffusion coefficient on angular momentum is ignored, and an averaged value is used, which may introduce 
systematic errors. One of our goals is to check the viability of such averaging.

The expressions for $D_{\alpha\beta,b}$ for the isothermal bulge are somewhat more complicated, so we do not list them 
here. (They can be found in \cite{a0307524}). 
The features of expressions are similar to (\ref{diff_coefs}), except that they are multiplied by $Q^{-2/3}$. 

Actually in the calculation we used the coefficients computed for the exact stellar distribution function, which in 
turn was derived from given density and potential using Eddington inversion formula \cite{BT}. 
These expressions are essentially interpolating between two limiting cases (actually, close to 
the minimum of two values ($D_{\alpha\beta,c}$ and $D_{\alpha\beta,b}$) for each coefficient); 
the transition occurs at $Q \sim 0.05$, which corresponds to spatial radius of roughly $0.2r_h$.

\subsection{Boundary conditions}

The diffusion coefficients tend to zero at $Q=0$ and $R=1$, so we do not need any boundary conditions there except 
regularity. The presence of black hole induces boundary condition at $R=R_g(Q)$, where $R_g=L_g^2/L_c^2$, 
$L_g=4GM_{bh}/c$ is the minimal angular momentum of a periodic orbit with $|E|\ll c^2$. 
The region $R<R_g$ is denoted as loss cone. 
We follow \citet{LS} in their treatment of the boundary condition, namely: there exist two absorption regimes -- 
empty loss cone, or random-walk capture (for $Q<Q_{cr}$), and full loss cone, or ``pinhole'' limit. 
The key difference between them is the ratio of $\Delta R$, root mean square change of $R$ during one orbital period, 
to $R_g$, the absorption boundary. 
In the former case, $\Delta R \ll R_g$; particles move in $R$-space by small steps in random direction (hence the 
name), and those getting into region $R<R_g$ are eliminated in one orbital time. To create a flux of particles into the 
black hole, it is therefore necessary to have gradient in distribution function, $\d f/\d R$.
In the opposite case, however, a particle that had $R<R_g$ at some point of trajectory, can easily diffuse out of the 
loss cone before it actually reaches its orbit pericenter and can be removed, and vice versa, particle that initially 
had $R \gg R_g$ (but $R\lesssim \Delta R$) can be captured by the end of its orbital period.
So the capture rate depends on amount of particles with small $R$, i.e. on value of $f$ near 0 rather than on its 
gradient. 

The boundary condition may be expressed as
\begin{equation}  \label{bound_cond}
\left. \left(f - \alpha R_g \frac{\d f}{\d R}\right) \right|_{R=R_g} = 0 \;,
\end{equation}
where $\alpha$ it taken from \citet{CK}:
\begin{subequations}  \label{alpha_bc}
\begin{eqnarray}
\alpha &=& \left\{ \begin{array}{lll} 0.824\sqrt{q}+0.186q &,& q< 1 \;;\\ q &,& q> 1 \;;\end{array}\right. \label{alpha}\\
q      &=& \left.\left(D_{RR}/R\right)\right|_{R\to 0} \,T_{orb}/R_g  \label{q}
\end{eqnarray}
\end{subequations}

The critical value of $Q$ separating the two regimes, for which $q=1$, is found to be within SMBH region of influence 
for present-day Milky Way parameters. As seen from (\ref{bound_cond}), in the pinhole regime the gradient of $f$ is 
smaller and hence the flux into SMBH is lower \citep{a0503712}.

\section{Solution of the diffusion equation}

\subsection{One-dimensional analysis}  \label{sec_onedim}

We begin our consideration of DM diffusion with analysis of one-dimensional limiting cases. A common approach is to
assume a logarithmic dependence of $f$ on $L$ (or $R$), which follows from steady-state solution of full 2-D equation
\citep{LS, CK}, and solve one-dimensional diffusion equation for energy, that includes loss term derived from the
solution for angular momentum. Actually, this is valid only for quasi-stationary solution for stars around black hole,
when the loss of stars in the BH is balanced by their influx from higher energies. 

In the case of dark matter, the situation is different, since there exist no steady state due to the fact that DM
particles are heated by stars and gain energy rather than lose, and do not provide the required influx.
So an accurate time-dependent solution of full 2-D equation is needed, at least in order to verify the validity of
reduction to one-dimensional diffusion for energy.

We first discuss the effect of one-dimensional diffusion for angular momentum in absence of diffusion for energy.
From (\ref{diff_coefs}c) it follows that $D_{RR} \approx D\,R$ for low $R$. The stationary solution of equation
$0=\frac{\d f}{\d t} = D\frac{\d}{\d R}\left(R\frac{\d f}{\d R}\right)$ has the form
\begin{equation}  \label{f_near_Rg}
f(R)=f_g \left(1+\frac{1}{\alpha} \ln\frac{R}{R_g}\right) \;,
\end{equation}
where $\alpha$ is defined in (\ref{alpha}). 
So for low $R$ the function has logarithmic form, as mentioned earlier, and the flux $S$ through absorption boundary
$S = D f_g/\alpha$ is balanced by flux from higher $R$. 

In the model B, $f(R) \propto R^{-\beta}$ at $t=0$, and for $t>0$ can be approximated as 
\begin{equation}
f(R,t) \approx \left\{ \begin{array}{lll} f_g(t) \left(1+\frac{1}{\alpha} \ln\frac{R}{R_g}\right) &,& R_g<R<R_1(t) \\
R^{-\beta} &,& R_1<R\leqslant 1 \end{array} \right.
\end{equation}
So for $R<R_1$ a logarithmic profile has been established, and outside $R_1$ the initial value is still conserved;
the two profiles coincide at $R=R_1$.
The values of $R_1$ and $f_g$ change in time so that the mass loss rate $\frac{d}{dt} \int_{R_g}^1 f(R,t)\,dR$ equals 
the flux to black hole $S$. This regime holds until $R_1$ reaches its maximum of unity (at time $t_1 \sim 1/D$);
then the logarithmic profile is established for all $R$, and its amplitude exponentially decreases with decay time 
$\tau=1/D \times(\ln(1/R_g)+\alpha)$. 

The above approximate arguments give correct qualitative results for one-dimensional diffusion along angular momentum. 
Global logarithmic profile is established at $t \gtrsim 1/D$ and decreases in amplitude with characteristic time
$\tau$ (which is larger for full loss cone, when $\alpha \gg 1$). The loss rate is nearly constant (actually,
logarithmically depends on $t$) for $t< 1/D$ and decreases exponentially for larger $t$, proportionally to the
amplitude of logarithmic profile.

\begin{table*}[t]
\caption{Variants of calculation. Initial halo models are described in Table~\ref{table_init_models}.
$M_{init}$ is the initial DM mass within $r_h=2$~pc, $M_{capt}$ -- amount of DM 
captured by SMBH (in parentheses -- from $r<r_h$), $M_{evap}$ -- DM mass evaporated from $r<r_h$, $M_{ann}$ -- 
annihilated mass (all of these -- for present time, $t=10^{10}$~yr, in $M_\odot$).
The final two columns give logarithm of astrophysical factor $J$ in annihilation flux, averaged over solid angle 
$10^{-3}$~sr (and $10^{-5}$~sr in parentheses), for $t=0$ and $t=t_{Hubble}$, respectively.\protect\\
All variants except 7 have $M_{bh}={\rm const}=3\times 10^6\,M_\odot$, in var.7 $M_{bh}$ rises from $3\times 10^4\,M_\odot$ 
to its present-day value as $\sqrt{t}$.
All variants except 6 are solved using two-dimensional equation; var.6 uses simplified one-dimensional diffusion 
equation for energy as described in Section~\ref{sec_onedim}.
In variants 2 and 5 maximal annihilation cross-section 
$\langle \sigma_a v_\chi\rangle = 3\times 10^{-26}\,\mbox{cm}^3/\mbox{s}$ and particle mass $m_\chi=50$~GeV was used, 
in others annihilation was not included.}  \label{table_variants}

\tabcolsep=2pt
\begin{tabular}{llll@{$\quad$}l@{$\quad$}l@{$\quad$}lll}
$\mbox{Var}$& \multicolumn{2}{l}{model, features}& $M_{init}$ & $M_{capt}\,(M_{capt}^{h})$ & $M_{evap}$ & $M_{ann}$ 
& $\log\overline{J_3} (\overline{J_5}), t$=0 & --''--, $t=t_{H\!ub}$\\
\hline
1 & B1 & Moore, reference var.     & $1.32\times 10^6$ & $1.78\times 10^5\; (1.00\times 10^5)$ & $5.89\times 10^5$ & --               & 15.61 (17.61) & 6.20 (8.19) \\
2 & B1 & maximal annihilation      & $1.32\times 10^6$ & $1.64\times 10^5\; (8.69\times 10^4)$ & $5.77\times 10^5$ & $3.04\times 10^4$& 15.61 (17.61) & 6.13 (8.12) \\
3 & B2 & radial anisotropy         & $9.79\times 10^5$ & $1.94\times 10^5\; (8.55\times 10^4)$ & $3.98\times 10^5$ & --               & 14.54 (16.54) & 5.93 (7.93) \\
4 & B3 & NFW, reference var.       & $3.05\times 10^5$ & $3.86\times 10^4\; (2.01\times 10^4)$ & $1.28\times 10^5$ & --               & 13.56 (15.56) & 5.05 (7.04) \\
5 & B3 & maximal annihilation      & $3.05\times 10^5$ & $3.73\times 10^4\; (1.88\times 10^4)$ & $1.27\times 10^5$ & $2.72\times 10^3$& 13.56 (15.56) & 5.00 (6.99) \\
6 & B3 & 1-D diffusion             & $3.05\times 10^5$ & $4.05\times 10^4\; (2.08\times 10^4)$ & $1.31\times 10^5$ & --               & 13.56 (15.56) & 5.03 (7.02) \\
7 & B3 & $M_{bh} \propto \sqrt{t}$ & $2.69\times 10^5$ & $2.17\times 10^4\; (8.85\times 10^3)$ & $1.36\times 10^5$ & --               & 14.22 (16.22) & 5.02 (7.01) \\
8 & B4 & Einasto profile           & $1.47\times 10^5$ & $1.77\times 10^4\; (8.49\times 10^3)$ & $5.74\times 10^4$ & --               & 12.13 (14.13) & 4.56 (6.54) \\
9 & A1 & Gurevich-Zybin profile    & $2.17\times 10^6$ & $6.75\times 10^5\; (2.68\times 10^5)$ & $7.22\times 10^5$ & --               & 12.26 (14.26) & 6.56 (8.56) \\
\hline
\end{tabular}
\end{table*}

We now turn to one-dimensional diffusion for energy in the absence of loss terms. 
First we consider the evolution of $f$ in the bulge outside SMBH domain of influence. In this region $D_{QQ} \propto Q^{4/3}$, 
the initial condition has the power-law form (\ref{f_init_QR}): $f(Q)\propto Q^{-n}$ ($\frac{1}{2}<n<\frac{2}{3}$), 
and the one-dimensional equation for diffusion along energy can be solved analytically.
The solution outside SMBH domain of influence may be approximated as
\begin{equation}
f(Q,t) \approx  \left\{ \begin{array}{lll}
 {\rm const}\times t^{-3/2n} &,& Q<Q_1(t) \;,\\
 f_0(Q)                  &,& Q>Q_1(t) \;.\end{array} \right.
\end{equation}
$Q_1(t)$ is defined to match these two asymptotes. This means that diffusion effectively blurs $f$ for low $Q$, 
driving it towards plateau with ever-diminishing amplitude (as a power-law of $t$). 

Then we consider the vicinity of SMBH, where $D_{QQ} \propto Q^2$. The equation then admits solution in the form 
$f(Q) \propto Q^{-n} \exp(-t/\tau) + \const$, where $\tau$ is the relaxation time (as said above, it is independent of 
energy in the case considered, but still depends on $R$: relaxation goes on faster for smaller angular momenta). If we 
neglect the dependence of $\tau$ on $R$, we may come to the following conclusion: as long as $t<\tau$, the evolution of 
$f$ (and hence density) is only in amplitude which drops exponentially, not the shape. When the exponent becomes 
significant (or $t>\tau$), the function tends to constant value which is set by the outer boundary condition 
(where it is determined by the solution outside SMBH domain of influence described above). Since the relaxation time 
in the inner area is always shorter than in bulge, this constant (with respect to $Q$) value is determined solely by 
the diffusion in bulge and is decreasing in time. The constant value of $f$ implies the density profile 
$\rho \propto r^{-3/2}$, as mentioned in \citep{a0308385}. 
(This is different in the case of stars, where the effect of dynamic friction should be included in the equation, 
and the corresponding solution has the Bahcall-Wolf density profile $\rho \propto r^{-7/4}$). However, the profile 
is not stationary and its amplitude decreases in time (as noted in \citep{a0311594}), which is attributed to ongoing 
heating of dark matter particles by stars, continued mainly outside $r_h$ after one relaxation time. 
The amplitude then decreases as power-law of time: $\rho \propto t^{-3/2\,n} = t^{-(6-\gamma)/(8-2\gamma)}$.

We want to draw attention to the importance of considering both regions (bulge and SMBH sphere of influence) in order 
to obtain the density profile and annihilation signal evolution for $t \sim t_{r, center}$. For both cases $t \gg t_r$ 
and $t \ll t_r$ analytical estimates can be obtained, but for transition period $t_r \lesssim t \lesssim 4t_r$ exact 
calculation is required. Specifically, Galactic center is just in this category.

Thus far we have considered only one-dimensional approximations, with the diffusion along angular momentum caused by 
capture of DM particles by the black hole, and diffusion along energy caused by heating on stars. 
A common approach to solution of the two-dimensional equation is to consider one-dimensional diffusion along energy 
with additional loss term due to capture of particles by black hole, caused by diffusion along momentum. 
The loss term is taken from the stationary solution of diffusion equation for momentum, and introduced into the 
equation for energy as an additional term $S$. 
The equation itself is written for averaged $\overline f(Q) = \int_0^1 f(Q,R)\, dR$. 
The magnitude of the loss term $S$ is taken from the boundary condition at the black hole: 
$S=D\,f_g/\alpha$, where $f_g$ is related to $\overline f$ by (\ref{f_near_Rg}). 
The diffusion coefficient $D_{EE}$ also should be averaged over $R$, despite the fact that it 
significantly varies with $R$ (see (\ref{diff_coefs}a)). 
The validity of this reduction should therefore be verified by comparison with solution of full 
two-dimensional equation, which was one of aims of the present study.

\subsection{Two-dimensional equation: variants of calculation}

In order to address different issues mentioned in Section~\ref{sec_overview}, we have performed several variants of 
calculation, which differ in initial conditions and in account for distinct factors of evolution. All of them are 
listed in Table~\ref{table_variants}.

The full two-dimensional diffusion equation (\ref{diff_eq}) was integrated on rectangular grid with variable space- 
and time-steps. The boundary condition at $R=R_g$ was treated in the same way as in \citep{CK} (section Vc), 
i.e. for $R_g$ less than $\Delta R_1$ (the size of first cell), flux $S$ through absorption boundary was considered 
as flux through $R=0$, which was related to the value of $f$ in the center of the cell via (\ref{f_near_Rg}) with 
$S=D f_g/\alpha$. The spatial density profile was recalculated every few timesteps and used to compute orbit-averaged 
annihilation rate at each $\{Q, R\}$ in variants 2 and 5. 

The variants are chosen to investigate the relative role of different factors. Five different initial models were 
chosen (Sec.~\ref{sec_initmodels}). The model B3 (NFW profile with isotropic velocity distribution) is a basic model 
to study the influence the one-dimensional approximation (Var.6) and growing BH mass (Var.7), which are to be 
compared with reference Var.4. From the existence of quasars at $z \approx 6$ \citep{a0512080} we know that SMBHs 
were already massive at very early times, but the situation is also possible when the black hole grows slowly from 
small initial seed due to accretion of stars and gas. The growth law is taken to be $M_{bh} \propto \sqrt{t}$ 
\citep{a0403023}; the SMBH radius of influence and the total gravitational potential are changed accordingly.
In other variants $M_{bh}$ was held constant, and gravitational potential remained fixed.

The influence of velocity anisotropy is studied in Var.3 for Moore profile (from the $\beta-\gamma$ relation 
of \citet{a0411473} it is naturally to expect more radial anisotropy for steeper profiles), 
which should be compared to Var.1. 

Additionally, annihilation was accounted for in Vars.2 and 5, with maximal annihilation cross-section of 
$\langle \sigma_a v_\chi\rangle = 3\times 10^{-26}\,\mbox{cm}^3/\mbox{s}$ and particle mass $m_\chi=50$~GeV, same as 
in \citep{a0501555}. These should be compared to Vars.1 and 4, respectively. In other variants annihilation was 
disregarded, but predictions of $\gamma$-ray flux were made (see below).

The prospects of indirect DM detection are related to registration of $\gamma$-rays from the DM annihilation 
in the most dense regions. One of such sites is the Galactic center, where indeed a $\gamma$-ray source was detected 
by several ground-based Cherenkov telescopes such as H.E.S.S. \citep{a0408145}, VERITAS \cite{a0403422}, 
CANGAROO \cite{a0403592} and MAGIC \cite{a0512469}, as well as space telescope EGRET \cite{a0210617}.

The $\gamma$-ray flux is usually split into two factors, one of them depending on particle physics parameters and 
the other ($J$ -- astrophysical factor) is determined by DM density distribution \citep{a9712318}.
\begin{equation}  \label{Phi_signal}
\Phi \simeq \Phi_0 \frac{\langle \sigma_a v_p\rangle}{3\times 10^{-26} \mbox{cm}^3/\mbox{s}} 
  \left(\frac{10\,\mbox{GeV}}{m_p}\right)^2  \overline{J_{\Delta\Omega}}\,,
\end{equation}
where $\Phi_0=5.6\times 10^{-8}\,\mbox{cm}^{-2}\mbox{s}^{-1}$, and $\overline{J_{\Delta\Omega}}$ is dimensionless 
integral of squared density, averaged over a solid angle $\Delta\Omega$:
\begin{equation}  \label{J_av}
\overline{J_{\Delta\Omega}} = K\,\frac{1}{\Delta\Omega}\int dl \int_0^{\sqrt{\Delta\Omega/\pi}} \rho^2(r) 2\pi\psi\,d\psi
\end{equation}
$K^{-1} = (8\mbox{ kpc})(0.3\mbox{ GeV/cm}^3)^2$, and integral is taken over line of sight and the solid angle 
corresponding to telescope resolution. (For simplicity, the point-spread function is taken as Heaviside step function).

We calculate values of $\overline{J_5}$ and $\overline{J_3}$ corresponding to $\Delta\Omega=10^{-5}$ and $10^{-3}$. 
The former is the approximate resolution of GLAST \citep{a0603540} and modern Cherenkov telescopes such as HESS, the 
latter is the resolution of EGRET \citep{a0210617}. The spatial resolution is 15 and 150 pc, correspondingly.
It is obvious that for a power-law density profile $\rho \propto r^{-\gamma}$ with $\gamma>1.5$ the main contribution 
in (\ref{J_av}) comes from the inner boundary, and for $\gamma<1.5$ the situation is inverse.
Since we have adiabatically compressed dark matter halo, its density slope $\gamma'$ after contraction is always 
greater than 1.5 ($\gamma'=(6-\gamma)/(4-\gamma)$, where $\gamma$ is initial slope). 
Therefore, the flux is determined by dark matter distribution in the vicinity of black hole, and the additional 
contribution from the dark matter in bulge ($r \gg r_h$) is negligible. 
This is in contrast with the conclusion of \citep{a0610425}, where the authors did not use adiabatically contracted 
dark matter profile and hence the contribution from bulge was significant because $\gamma<1.5$.

Another observational constraint on dark matter distribution may come from the study of stellar orbits which 
measure the enclosed mass and may help to determine the fraction of dark mass near the SMBH \citep{a0602400}.

\begin{figure} 
$$\includegraphics{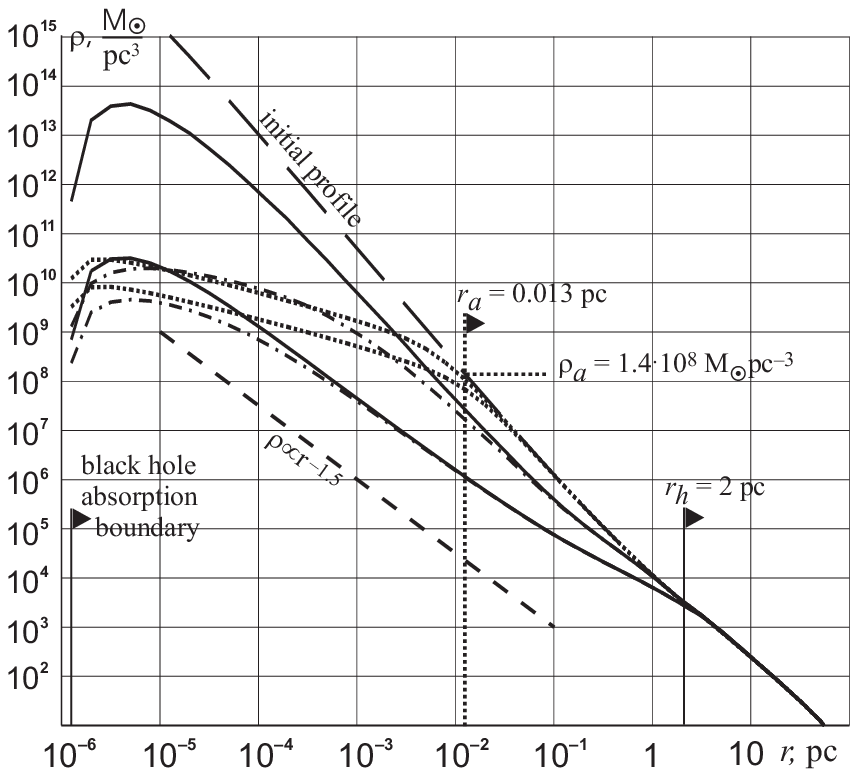} $$
\caption{
Density profiles for variants 4 (without annihilation, solid), 5 (with maximal annihilation, dot-dashed), 
and for pure annihilation without diffusion (dotted lines).
Top curve in each series is for $t=2.5$~Gyr, approximately one relaxation time; bottom is for $t=10$~Gyr. 
Initial adiabatically compressed profile is shown by long-dashed line. \protect\\
Without diffusion, the density profile would be a broken power-law with $\rho \propto r^{-0.5}$ for $r<r_a$, 
and $\rho \propto r^{-2.33}$ for $r>r_a$ (initial profile); 
$r_a$ is the annihilation radius at which the initial density equals $\rho_a$ (\ref{rho_a}) which is set at 
$1.4\times 10^8\,M_\odot\,\mbox{pc}^{-3}$ by now. \protect\\
Without annihilation the density would drop several times by one relaxation time in the whole region $r<r_h$ 
(top solid line), and drop dramatically by now (bottom solid line); since it becomes shallower than 
$\rho \propto r^{-1.5}$ (which is sketched as dashed line) inside $r_h$, the main contribution to annihilation 
factor $J$ comes from $r\sim r_h$. \protect\\
Combining together the effects of annihilation and diffusion (dot-dashed lines), we get profiles that are 
similar to pure diffusional at large radii and are cut by annihilation at very small radii. The $J$ factor 
by now is essentially the same as without annihilation and is not affected by this cutoff, but for $t=t_r$ the 
difference in $J$ between Var.4 and Var.5 is still large (see Fig.\ref{fig_J}).
} \label{fig_dens_annihil}
\end{figure} 

\begin{figure} 
$$\includegraphics{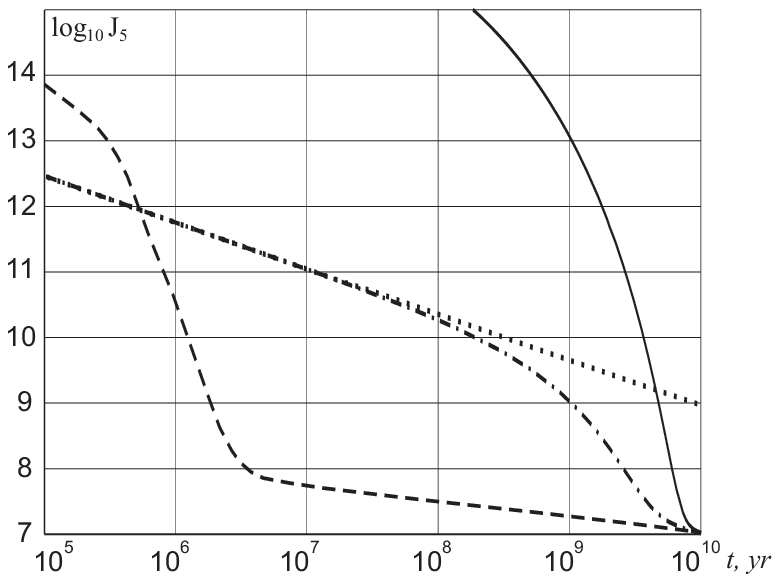} $$
\caption{
Evolution of annihilation factor $J_5$ for models 4 (without annihilation, solid), 5 (maximal annihilation, 
dot-dashed), 7 ($M_{bh}\propto \sqrt{t}$, dashed), and for pure annihilation without diffusion (dotted lines). \protect\\
The latter one is a power-law in time ($J \propto t^{-5/7}$); the case without annihilation has a sharp decline 
at $t \gtrsim t_r \approx 2.5$~Gyr which corresponds to transition from a steeper than $r^{-1.5}$ slope before 
relaxation to a shallower slope when the value of $J$ is dominated by density at $r\sim r_h$ and changes very 
weakly in time since $t\gtrsim 4t_r$. \protect\\
The model 5 with both annihilation and diffusion included is roughly an interpolation between these two extremes: 
before relaxation the value of $J$ is dominated by density at small radii ($\sim r_a$), and after it 
is identical to the model 4 without annihilation. The time after which values of $J$ for models 4 and 5 start to 
be equal happens to coincide with current time, which is by no means an rigorous result. \protect\\
The model 7 has much shorter relaxation time so the sharp decline occurs earlier, and since then the evolution 
is close to self-similar with very slow decline of $J$ in time. 
} \label{fig_J}
\end{figure} 

\begin{figure}[t] 
$$\includegraphics{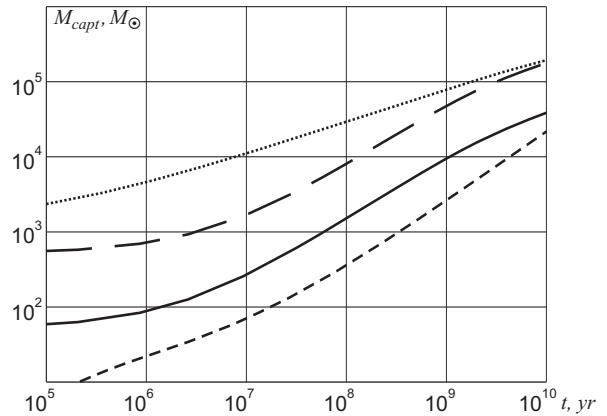} $$
\caption{
Evolution of dark matter mass captured by black hole, for models 3 (dotted), 1 (long-dashed), 4 (solid) and 7 
(short-dashed line), from top to bottom. 
Models 1 (Moore profile) and 4 (NFW profile) differ in initial dark matter mass and hence the same difference in 
captured mass, with similar dynamics. 
Model 3 is Moore profile with initial radial anisotropy: initially the captured mass grows faster because of 
greater abundance of particles with low angular momenta; however, after one relaxation time the difference in 
initial conditions almost disappears and the capture rate and total mass becomes similar.
Model 7 features small initial seed black hole mass and shorter relaxation timescale, hence the capture rate 
is lower and heating goes on faster; by now the total captured mass is smaller.
} \label{fig_Mcapt}
\end{figure} 

The results of calculation are presented in Table~\ref{table_variants}. 
The five initial models differ in normalization, i.e. in the initial dark matter mass $M_{init}$ within $r_h$. 
Since the dynamics of dark matter is governed by linear equations (except for annihilation), the results may well be 
scaled to match each other. The similarity between different models then becomes apparent. 
Approximately 10--15\% of $M_{init}$ is captured by black hole by $t=10^{10}$~yr ($M_{capt}$); 
half of this mass comes from the SMBH sphere of influence ($r<r_h, M_{capt}^h$). About 40\% of $M_{init}$ has been 
evaporated from the central region.

The difference between variants with the same initial model are significant only for $t\lesssim 4t_r = 10^{10}$~yr, 
which is by coincidence the total time of evolution. $t_r$ (\ref{t_relax}) is the relaxation time inside $r_h$ and 
is independent of radius. So the current state (and further evolution) is basically independent of initial conditions, 
although if $t_r$ was, say, twice as large, then the difference would be much more prominent by present time. 

First we consider basic variant 4 without annihilation.
Since the diffusion along energy goes on faster for low angular momenta, the particles with both low energy and low 
angular momenta are quickly heated up or captured by black hole, therefore the density in the immediate vicinity 
of the black hole quickly drops. 
However, by one relaxation time it decreases roughly by the same factor of few in the whole region 
$r \lesssim 0.2r_h$ (Fig.~\ref{fig_dens_annihil}, upper solid), except for the very inner radii. 
By this time the density profile is still steeper than $r^{-1.5}$ inside $r_h$, and hence the integral 
$\int \rho^2(r) r^2\,dr$ which determines the $\overline J$-factor in (\ref{J_av}) depends on the low-$r$ density 
behavior. 
Then the heating starts to dominate, and within few $t_r$ the distribution function and density tend to nearly 
universal form: for $r\lesssim 0.1r_h$ density $\rho \sim r^{-1.5}$, near $r \sim r_h$ density profile is a little bit
shallower, and further away it is identical to initial profile. So most of $\overline J$ is gained at $r \sim r_h$, 
and hence the difference between variants $4-7$ almost vanishes. The density profile inside $r_h$ then drops nearly 
self-similarly (if we continue integration to $10^{11}$~yrs, density is reduced about an order of magnitude 
everywhere at $r<r_h$ while retaining its form).
It is important to note that after this regime has been established, diffusion {\sl outside} SMBH sphere of influence 
plays the key role.

In the models with initial dominance of low angular momentum orbits (B2 and A1) the initial flux of dark matter 
into the SMBH is larger, but then as the $R$-diffusion establishes near-logarithmic profile (at $t\gtrsim t_r$), 
the memory of initial conditions is erased. 

\begin{figure}[t] 
$$\includegraphics{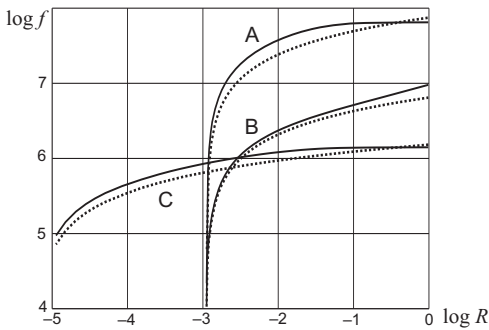} $$
\caption{ Distribution function $f(R)$ for fixed Q: solid lines -- Var.4, dotted lines -- Var.6 (1-D treatment).
Curves A and C are for $Q=10^{-5}$ (corresponding to circular orbit with $r_{circ}=10^{-3}r_h$, $R_g=0.0011$, and 
$\alpha=1.6\times 10^{-3}$ (Eq.\ref{alpha_bc})), curve B is for $Q=10^{-3}$ ($r_{circ}=0.1r_h$, $R_g=1.1\times 10^{-5}$, 
and $\alpha=0.56$). Curves A and C are for $t=0.01 t_{r}$, curve B is for $t=t_{r}$.\protect\\
The difference between solution of exact two-dimensional equation and simplified one-dimensional treatment, 
where profile is exactly logarithmic (Eq.\ref{f_near_Rg}), is rather moderate: 
for early times (curves A and B) the logarithmic profile in the exact solution is not yet established for all $R$, 
and for high $R$ the function is close to initial constant value; while for intermediate times ($t \sim t_r$, curve B) 
the difference is due to intrinsically two-dimensional character of evolution. (For $t=0.1 t_r$ or $t=10t_r$ the 
corresponding curves would be very close). In all cases, the behavior of $f(R)$ near loss-cone boundary $R_g$ is very 
close to logarithmic; the difference from 1-D case is only in amplitude. \protect\\
The curves A and B illustrate random-walk regime, or empty loss cone; curve C suggests that in pinhole regime 
($\alpha \gg 1$) $f(R)$ is approximately constant.
} \label{fig_fR}
\end{figure} 

\begin{figure}[t] 
$$\includegraphics{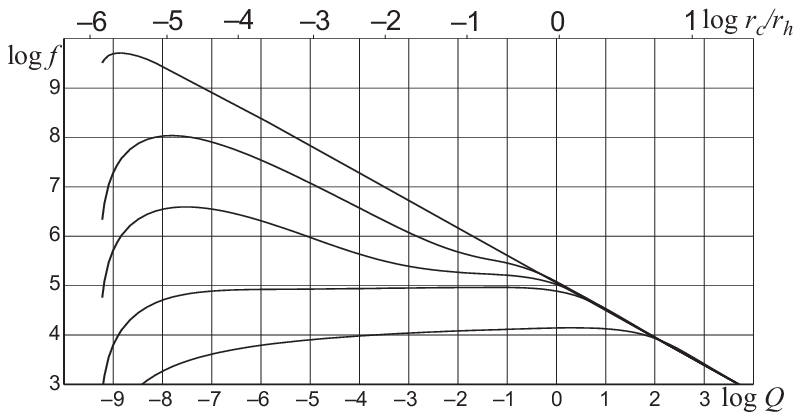} $$
\caption{
Evolution of $R$-averaged distribution function in Var.4 for different $t$ (from top to bottom): 
$t=0$, $t=2.5\times 10^9 \mbox{ yr}\approx t_r$, $t=5\times 10^9$, $t=10^{10}$ and $t=10^{11}$~yr.
Initially $f\propto Q^{-5/9}$; by one relaxation time $f$ drops by an order of magnitude nearly uniformly 
inside SMBH sphere of influence, where $t_r$ is constant; by $4t_r$ it is almost flat for $Q<1$ and continues 
to drop as $t^{-1}$, as described in Section~\ref{sec_onedim}. 
The slope $Q^{-5/9}$ corresponds to density profile $\rho \propto r^{-7/3}$ inside $r_h$ and $r^{-5/3}$ outside;
flat distribution -- to $r^{-3/2}$ and $r^0$, correspondingly.
The top axis shows $r_c$ -- radii of circular orbits for given $Q$, in units of $r_h$.
} \label{fig_fQ}
\end{figure} 

The inclusion of self-annihilation (in Vars.~2 and 5) affects only the inner region in the manner that the density 
does not exceed the ``annihilation weak cusp'' (not plateau). Consequently, the $\overline J$ value diminishes faster 
than without annihilation (Fig.~\ref{fig_J}), but after $t\gtrsim t_r$ the heating starts to prevail over annihilation 
even in the very inner region, and everything again goes as in basic variant.

The variant 7 which features a black hole growing from small initial seed differs in the way that relaxation time 
inside $r_h$ is less and a nearly self-similar relaxed solution is established earlier. In this case the total dark 
matter mass captured by the black hole is also less than in the case of constant black hole mass (Fig.~\ref{fig_Mcapt}).
In all variants the total captured mass is much less than $M_{bh}$, so the neglect of the absorption on the black hole 
mass is justified.

Finally, comparison of one-dimensional solution (Var.6) to the two-dimensional one (Var.4) reveals very little 
difference in either final result or dynamics. This is because $R$-profile of distribution function at fixed $Q$ 
is reasonably well described by a logarithm, failing only at very early ($t\ll t_r$) or intermediate ($t\sim t_r$) 
times (Fig.~\ref{fig_fR}). 
Therefore, the commonly used approach of reduction to one-dimensional diffusion equation is a good approximation, 
at least for the class of initial distribution functions that have weak dependency on $R$ 
(the case of velocity isotropy -- model B3 -- being in this class). 
Moreover, the evolution of $R$-averaged distribution function is qualitatively consistent with simple analytical 
description of Section~\ref{sec_onedim} (Fig.~\ref{fig_fQ}).

\section{Discussion and conclusions}

We have considered the evolution of dark matter distribution in the Galactic center caused by gravitational 
scattering on stars of galactic bulge, absorption by the supermassive black hole and annihilation. 
The spherically symmetric evolution is described by two-dimensional Fokker-Planck equation in phase space of 
energy and angular momentum. One of main goals was to investigate how the difference in initial conditions and 
in various dynamical factors affects the final result.

\enlargethispage{2\baselineskip}

It turns out that at least for our Galaxy there has been enough time to eliminate most of the differences and to 
establish a nearly universal distribution and density profile with the following features: for $r\lesssim 0.1\;r_h 
\approx 0.2$~pc the density is close to power-law $\rho\propto r^{-\gamma}$ with slope $\gamma \approx 1.5$ or less, 
farther out the slope becomes shallower, and even farther it joins the initial profile which is already steeper than 
$r^{-1.5}$ because of adiabatic compression in the bulge. 
Therefore, as regarding the possible annihilation signal, after approximately four relaxation times 
it is determined by DM distribution near $r_h$ and depends only on initial profile normalization, 
not the details of evolution. 
This explains the low scatter in $\overline J$ values in the last column of Table~\ref{table_variants}, 
the most probable range for $\overline {J_5}$ being $10^6 - 10^8$. (This is in moderate agreement with results of 
\citet{a0504422}, though in our calculations the difference between models with no annihilation and maximal 
annihilation is much smaller). 
The low scatter in $J$ means that constraints on the microphysical nature of dark matter following from 
Eq.(\ref{Phi_signal}) will be more robust, if the $\gamma$-ray flux detected from the Galactic center is at least 
partially attributed to dark matter annihilation \citep{a0610509, a0603540}.

The different initial models also demonstrate similar proportions of dark matter mass that has been 
captured by the black hole ($\sim 10-15\%$) and evaporated from the central 2 pc ($\sim 40\%$ of the initial DM mass 
within 2 pc).
A number of previous studies suggested that the DM may constitute a significant fraction of the total SMBH mass. 
However, in these papers either the loss cone was assumed to be always full \cite{a0112096, a0201153}, or the 
diffusion along energy axis (heating) was disregarded \cite{a0306490, a0307524}, which led to higher estimate of the 
captured mass compared to the present study. It remains in question whether a non-spherically-symmetric initial 
distribution could sustain a full ``loss cone'' due to chaotic randomization of orbits \cite{a0112096, a0302296}.
This issue will be addressed in a future paper.

We found that the one-dimensional diffusion approximation works good for models with $f$ initially independent 
on angular momentum.

Comparison with paper \cite{a0311594} where 1-D equation was solved shows qualitatively similar behavior 
of $R$-averaged distribution function and density evolution: for $t\lesssim t_r$ the density drops self-similarly 
in the SMBH region of influence, and for $t\gg t_r$ it tends to $r^{-3/2}$ form. The quantitative difference 
might be attributed to different stellar density profile outside $\sim 0.1r_h$: the evolution in the transition 
spatial ($r\sim r_h$) and temporal ($t \sim t_r$) regions essentially depends on the magnitude of diffusion 
coefficients which are determined by stellar density.

There are other targets of indirect dark matter search, except the Galactic center, such as nearby galaxies or 
globular clusters \citep{a08011708}. 
The importance of the processes discussed in this paper on the $\gamma$-ray flux from these other 
targets depends on whether they are dynamically old, i.e. is their relaxation time in the very center less than 
Hubble time. For example, dark matter dominated dwarf spheroidal galaxies such as Draco have relaxation times well 
exceeding $10^{10}$~yr, so the scattering off by stars is unimportant, and the flux is determined by initial 
conditions and annihilation cross-section \citep{a0701426}. 
On the other hand, globular clusters such as G1 in Andromeda Galaxy are dynamically old, and are expected 
to be on the relaxed stage of evolution and therefore have less scatter in $J$, 
hence they are also promising in constraining dark matter parameters \citep{a07124181}.
(However, in this case one should account for evolution of stellar distribution as well). 
Finally, galaxies on the intermediate stage of evolution such as M32 which has central relaxation time of order 
$2-3\times 10^9$~yr \citep{a9806277} may exhibit very different values of $J$ (a drop by many orders of magnitude 
during several $t_r$ is seen on Fig.~\ref{fig_J}).

The concentration of dark matter in the center of our Galaxy is a promising target of indirect dark matter detection, 
and its dynamical evolution is crucial for constraining dark matter parameters.

The authors thank V.~L.~Ginzburg for countenance of this research and A.~Ilyin, V.~Sirota and K.~Zybin for 
fruitful discussions. 
This work was supported by Russian Fund for Basic Research (grant No. 07-02-01128-a)
and President's grant council (grants No. NSh-438.2008.2 and MK-446.2008.2).



\begin{thebibliography}{99}

\bibitem[\protect\citeauthoryear{Silk}{2002}]                   {a0110404} J.~Silk, Int. J. Mod. Phys. A {\bf 17} Suppl.1, 167 (2002).
\bibitem[\protect\citeauthoryear{Bertone, Hooper \& Silk}{Bertone et al.}{2004}]        {h0404175} G.~Bertone, D.~Hooper, and J.~Silk,  Phys. Rept. {\bf 405}, 279 (2005).
\bibitem[\protect\citeauthoryear{Springel et al.}{2005}]        {a0504097} V.~Springel, S.~White, A.~Jenkins et al., Nature {\bf 435}, 629 (2005).
\bibitem[\protect\citeauthoryear{Blumenthal et al.}{1986}]      {Blum} G.~Blumenthal, S.~Faber, R.~Flores, and J.~Primack,  Astrophys. J. {\bf 301}, 27 (1986).
\bibitem[\protect\citeauthoryear{Gnedin et al.}{2004}]          {a0406247} O.~Gnedin, A.~Kravtsov, A.Klypin, and D.~Nagai,  Astrophys. J. {\bf 616}, 16 (2004).
\bibitem[\protect\citeauthoryear{Sellwood \& McGaugh}{2005}]    {a0507589} J.~Sellwood and S.~McGaugh,  Astrophys. J. {\bf 634}, 70 (2005). 
\bibitem[\protect\citeauthoryear{Vasiliev}{2006}]               {a0601669} E.~Vasiliev, JETP Lett. {\bf 84}, 2 (2006).
\bibitem[\protect\citeauthoryear{Ilyin, Zybin \& Gurevich}{Ilyin et al.}{2004}] {a0306490} A.~Ilyin, K.~Zybin, and A.~Gurevich; J. Exp. Theor. Phys. {\bf 98}, 1 (2004).
\bibitem[\protect\citeauthoryear{Zelnikov \& Vasiliev}{2005a}]  {a0307524} M.~Zelnikov and E.~Vasiliev, Int. J. Mod. Phys. A {\bf 20}, 4217 (2005).
\bibitem[\protect\citeauthoryear{Bertone \& Merritt}{2005a}]    {a0501555} G.~Bertone and D.~Merritt,  Phys. Rev. D {\bf 72}, 103502 (2005). 
\bibitem[\protect\citeauthoryear{Zeldovich et al.}{1980}]	{Zeld} Ya.~Zeldovich, A.~Klypin, M.~Khlopov, and V.~Chechetkin, Sov. J. Nucl. Phys. {\bf 31}, 664 (1980).
\bibitem[\protect\citeauthoryear{Gondolo \& Silk}{1999}]        {a9906391} P.~Gondolo and J.~Silk,  Phys. Rev. Lett. {\bf 83}, 1719 (1999).
\bibitem[\protect\citeauthoryear{Bertone \& Merritt}{2005b}]    {a0504422} G.~Bertone and D.~Merritt,  Mod. Phys. Lett. A{\bf 20}, 1021 (2005).
\bibitem[\protect\citeauthoryear{Mambrini et al.}{2006}]        {h0506204} Y.~Mambrini, C.~Munos, E.~Nezri et al.,  JCAP {\bf 0601}, 010 (2006).
\bibitem[\protect\citeauthoryear{Gurevich \& Zybin}{1995}]      {GZ} A.~Gurevich and K.~Zybin, Phys.-Usp. {\bf 38}, 687 (1995).
\bibitem[\protect\citeauthoryear{Nusser}{2001}]                 {a0008217} A.~Nusser,  Mon. Not. R. Astron. Soc. {\bf 325}, 1397 (2001).
\bibitem[\protect\citeauthoryear{Mamon \& {\L}okas}{2005}]      {a0405491} G.~Mamon and E.~{\L}okas,  Mon. Not. R. Astron. Soc. {\bf 363}, 705 (2005).
\bibitem[\protect\citeauthoryear{Austin et al.}{2005}]          {a0506571} C.~Austin, L.~Williams, E.~Barnes et al., Astrophys. J. {\bf 634}, 756 (2005).
\bibitem[\protect\citeauthoryear{Navarro, Frenk \& White}{Navarro et al.}{1997}]        {a9611107} J.~Navarro, C.~Frenk, and S.~White,  Astrophys. J. {\bf 490}, 493 (1997).
\bibitem[\protect\citeauthoryear{Moore et al.}{1999}]           {a9903164} B.~Moore, T.~Quinn, F.~Governato et al.,  Mon. Not. R. Astron. Soc. {\bf 310},  1147 (1999).
\bibitem[\protect\citeauthoryear{Navarro et al.}{2004}]         {a0311231} J.~Navarro, E.~Hayashi, C.~Power et al.,  Mon. Not. R. Astron. Soc. {\bf 349}, 1039 (2004).
\bibitem[\protect\citeauthoryear{Merritt et al.}{2005}]         {a0502515} D.~Merritt, J.~Navarro, A.~Ludlow, and A.~Jenkins, Astrophys. J. {\bf 624}, L85 (2005).
\bibitem[\protect\citeauthoryear{Gentile et al.}{2005}]         {a0506538} G.~Gentile, A.~Burkert, P.~Salucci, U.~Klein, and F.~Walter, Astrophys. J. {\bf 634}, L145 (2005).
\bibitem[\protect\citeauthoryear{Diemand, Kuhlen \& Madau}{Diemand et al.}{2007}]	{a0611370} J.~Diemand, M.~Kuhlen, and P.~Madau, Astrophys. J. {\bf 657}, 262 (2007).
\bibitem[\protect\citeauthoryear{Ferrarese \& Ford}{2005}]      {a0411247} L.~Ferrarese and H.~Ford, Space Science Reviews {\bf 116}, 523 (2005).
\bibitem[\protect\citeauthoryear{Ullio, Zhao \& Kamionkowski}{Ullio et al.}{2001}]      {a0101481} P.~Ullio, H.~Zhao, and M.~Kamionkowski,  Phys. Rev. D {\bf 64}, 043504 (2001).
\bibitem[\protect\citeauthoryear{Merritt et al.}{2002}]         {a0201376} D.~Merritt, M.~Milosavljevi\'c, L.~Verde, and R.~Jimenez,  Phys. Rev. Lett. {\bf 88}, 191301 (2002).
\bibitem[\protect\citeauthoryear{Vasiliev}{2007}]               {a07073334} E.~Vasiliev, Phys. Rev. D {\bf 76}, 103532 (2007).
\bibitem[\protect\citeauthoryear{Bahcall \& Wolf}{1976}]        {BW} J.~Bahcall and R.~Wolf, Astrophys. J. {\bf 209}, 214 (1976).
\bibitem[\protect\citeauthoryear{Lightman \& Shapiro}{1977}]    {LS} A.~Lightman and S.~Shapiro,  Astrophys. J. {\bf 211}, 244 (1977).
\bibitem[\protect\citeauthoryear{Cohn \& Kulsrud}{1978}]        {CK} H.~Cohn and R.~Kulsrud,  Astrophys. J. {\bf 226}, 1087 (1978).
\bibitem[\protect\citeauthoryear{Genzel et al.}{2003}]          {a0305423} R.~Genzel, R.~Sch\"odel, T.~Ott et al., Astrophys. J. {\bf 594}, 812 (2003).
\bibitem[\protect\citeauthoryear{Ghez et al.}{2005}]            {a0306130} A.~Ghez, S.~Salim, D.~Hornstein et al., Astrophys. J. {\bf 620}, 744 (2005).
\bibitem[\protect\citeauthoryear{Sch\"odel et al.}{2007}]       {a0703178} R.~Sch\"odel et al., Astron. Astrophys. {\bf 469}, 125 (2007).
\bibitem[\protect\citeauthoryear{Binney \& Tremaine}{1987}]     {BT} J.~Binney and S.~Tremaine ``Galactic dynamics'',  Princeton University Press (1987).
\bibitem[\protect\citeauthoryear{Eisenhauer et al.}{2005}]      {a0502129} F.~Eisenhauer, R.~Genzel, T.~Alexander et al., Astrophys. J. {\bf 628} 246 (2005).
\bibitem[\protect\citeauthoryear{Gnedin \& Primack}{2004}]      {a0308385} O.~Gnedin and J.~Primack,  Phys. Rev. Lett. {\bf 93}, 061302 (2004).
\bibitem[\protect\citeauthoryear{Merritt}{2004}]                {a0311594} D.~Merritt,  Phys. Rev. Lett. {\bf 92}, 201304 (2004).
\bibitem[\protect\citeauthoryear{Merritt, Harfst \& Bertone}{Merritt et al.}{2007}]     {a0610425} D.~Merritt, S.~Harfst, and G.~Bertone, Phys. Rev. D {\bf 75} 043517 (2007).
\bibitem[\protect\citeauthoryear{An \& Evans}{2006}]            {a0511686} J.~An and N.~Evans,  Astrophys. J. {\bf 642}, 752 (2006).
\bibitem[\protect\citeauthoryear{Hansen \& Moore}{2006}]        {a0411473} S.~Hansen and B.~Moore,  New Astron. {\bf 11}, 333 (2006).
\bibitem[\protect\citeauthoryear{Tremaine et al.}{2002}]        {a0203468} S.~Tremaine, K.~Gebhardt, R.~Bender et al., Astrophys. J. {\bf 574}, 740 (2002).
\bibitem[\protect\citeauthoryear{Zelnikov \& Vasiliev}{2005b}]  {a0503712} M.~Zelnikov and E.~Vasiliev, JETP Lett. {\bf 81}, 85 (2005).
\bibitem[\protect\citeauthoryear{Fan et al.}{2006}]		{a0512080} X.~Fan, M.~Strauss, G.~Richards et al., Astron. J. {\bf 131}, 1203 (2006).
\bibitem[\protect\citeauthoryear{Sirota et al.}{2005}]		{a0403023} V.~Sirota, A.~Ilyin, K.~Zybin, and A.~Gurevich, JETP {\bf 100}, 294 (2005).
\bibitem[\protect\citeauthoryear{Aharonian et al.}{2004}]	{a0408145} F.~Aharonian et al., Astron. Astrophys. {\bf 425}, L13 (2004).
\bibitem[\protect\citeauthoryear{Kosack et al.}{2004}]		{a0403422} K.~Kosack et al., Astrophys.J. {\bf 608} L97 (2004).
\bibitem[\protect\citeauthoryear{Tsuchiya et al.}{2004}]	{a0403592} K.~Tsuchiya et al., Astrophys. J. {\bf 606}, L115 (2004).
\bibitem[\protect\citeauthoryear{Albert et al.}{2004}]		{a0512469} J.~Albert et al., Astrophys. J. {\bf 638}, L101 (2006).
\bibitem[\protect\citeauthoryear{Hooper \& Dignus}{2004}]	{a0210617} D.~Hooper, B.~Dignus, Phys. Rev. D {\bf 70}, 113007 (2004).
\bibitem[\protect\citeauthoryear{Bergstr\"om, Ullio \& Buckley}{Bergstr\"om et al.}{1998}]	{a9712318}  L.~Bergstr\"om, P.~Ullio, and J.~Buckley, Astropart. Phys. {\bf 9}, 137 (1998).
\bibitem[\protect\citeauthoryear{Zaharijas \& Hooper}{2006}]	{a0603540} G.~Zaharijas and D.~Hooper, Phys. Rev. D {\bf 73}, 103501 (2006).
\bibitem[\protect\citeauthoryear{Hall \& Gondolo}{2006}]        {a0602400} J.~Hall and P.~Gondolo, Phys. Rev. D {\bf 74}, 063511 (2006).
\bibitem[\protect\citeauthoryear{Aharonian et al.}{2006}]	{a0610509} F.~Aharonian et al., Phys. Rev. Lett. {\bf 97}, 221102 (2006).
\bibitem[\protect\citeauthoryear{Zhao et al.}{2002}]        	{a0112096} H.~Zhao, M.~Haehnelt and M.~Rees, New Astron. {\bf 7}, 385 (2002).
\bibitem[\protect\citeauthoryear{MacMillan \& Henriksen}{2002}] {a0201153} J.~MacMillan and R.~Henriksen, Astrophys. J. {\bf 569}, 83 (2002).
\bibitem[\protect\citeauthoryear{Merritt \& Poon}{2004}]        {a0302296} D.~Merritt and M.~Poon, Astrophys. J. {\bf 606}, 788 (2004).
\bibitem[\protect\citeauthoryear{Wood et al.}{2008}]		{a08011708} M.~Wood, G.~Blaylock, S.~Bradbury et al., Astrophys. J. {\bf 678}, 594 (2008).
\bibitem[\protect\citeauthoryear{Sanchez-Conde et al.}{2004}]	{a0701426} M.~Sanchez-Conde, F.~Prada, E.~{\L}okas et al., Phys. Rev. D {\bf 76}, 123509 (2007).
\bibitem[\protect\citeauthoryear{Zaharijas}{2007}]              {a07124181} G.~Zaharijas, Phys. Rev. D, {\bf 78}, 027301 (2008).
\bibitem[\protect\citeauthoryear{Lauer et al.}{1998}]		{a9806277}  T.~Lauer, S.~Faber, E.~Ajhar, C.~Grillmair, and P.~Scowen,  Astron. J. {\bf 116}, 2263 (1998).
\end{thebibliography}
\end{document}